\documentclass[pra,superscriptaddress,showpacs,twocolumn]{revtex4-1}

\usepackage{amsmath,amssymb}
\usepackage{amsfonts}
\usepackage{graphics}
\usepackage{graphicx}

\newcommand{\bvec}[1]{{\mathbf #1}}

\newcommand{\beq}{\begin{eqnarray}}
\newcommand{\eeq}{\end{eqnarray}}

\usepackage{color}

\begin{document}

\title{Interacting quantum Hall states in a finite graphene flake and at finite temperature}

\author{Hank Chen}

\affiliation{Department of Mathematics, University of Waterloo, Waterloo, Ontario, N2L 3G1, Canada}

\author{Matthew R. C.  Fitzpatrick}

\affiliation{Department of Physics, Simon Fraser University, 8888 University Drive, Burnaby, British Columbia, V5A 1S6, Canada}

\author{Sujit Narayanan}

\affiliation{Department of Physics, Simon Fraser University, 8888 University Drive, Burnaby, British Columbia, V5A 1S6, Canada}

\author{Bitan Roy}
\affiliation{Department of Physics, Lehigh University, Bethlehem, Pennslyvania 18015, USA}

\author{Malcolm P. Kennett}

\affiliation{Department of Physics, Simon Fraser University, 8888 University Drive, Burnaby, British Columbia, V5A 1S6, Canada}

\date{\today}

\begin{abstract}
The integer quantum Hall states at fillings $\nu = 0$ and $|\nu| = 1$ in monolayer graphene have drawn much attention as they are generated by electron-electron interactions. 
Here we explore aspects of the $\nu = 0$ 
and $|\nu| = 1$ quantum Hall states relevant for experimental samples.  In particular, we study the effects of finite 
extent and finite temperature on the $\nu = 0$ state and finite temperature for the $\nu = 1$ state.  
For the $\nu = 0$ state we consider the situation in which the bulk is a canted antiferromagnet and use parameters 
consistent with measurements of the bulk gap to study the edge states in tilted magnetic fields in order to compare with
experiment [A. F. Young {\it et al.}, Nature {\bf 505}, 528 (2014)]. When spatial modulation of the 
order parameters is taken into account, we find that for graphene placed on boron nitride, the gap at the edge closes for magnetic fields comparable 
to those in experiment, giving rise to edge conduction with $G \sim 2e^2/h$ while the bulk gap remains almost unchanged.  
We also study the transition into the ordered state at finite temperature and field. 
We determine the scaling of critical temperatures as a function of 
magnetic field, $B$, and distance to the zero field critical point and find sublinear scaling with magnetic field for 
weak and intermediate strength interactions, and $\sqrt{B}$ scaling at the coupling associated with the zero field quantum critical point.
We also predict that critical temperatures for $\nu = 0$ states should be an order of magnitude higher than those for $|\nu| = 1$ states,
consistent with the fact that the low temperature gap for $\nu = 0$ is roughly an order of magnitude larger
than that for $|\nu| = 1$.

\end{abstract}

\maketitle

\section{Introduction}
The quantum Hall states in monolayer graphene reflect the Dirac nature of the low energy
quasiparticles, exhibiting 
plateaux for $\nu=\pm (4 n+2)$ at weak magnetic fields \cite{qhe-graphene-1,qhe-graphene-2}.
In a non-interacting picture the positions of these plateaux can be understood as arising from 
fourfold valley and spin degeneracy of two dimensional Dirac fermions \cite{sharapov}.
At stronger magnetic fields, additional plateaux arise at $\nu=0, \pm 1$ and $\pm 4$ \cite{Zhang2007}.
The $\nu = 0$ quantum Hall state in particular has attracted much recent experimental
\cite{Zhang2007,Andrei2010,YacobyPRB2013,NovoselovPNAS,Young2012,Andrei2009,Dean2011,Yacoby2012,Li2019,Hong2019} 
and theoretical \cite{Roy2014,Khveshchenko2001,HerbutPRL2006,HJR2009,herbut-qhe,herbut2007,KunYang2007,semenoff-zhou,barlas-review,kharitonov,Narayanan2018,goerbig-review} 
attention as it is an example of an integer quantum Hall state that is generated by electron-electron interactions.

In a strong magnetic field, electron-electron interactions are enhanced as kinetic energy is quenched by 
the formation of Landau levels (LLs) which can 
lead to the formation of ordered phases even for infinitesimally 
small interactions in the presence of a magnetic field, not only by splitting the half-filled zeroth LL (ZLL),
but also by simultaneously lowering the energies of all filled LLs with negative energies.
This phenomenon is known as magnetic catalysis \cite{Roy2014,herbut-qhe,herbut2007,roy-scaling,roy-inhomogeneous-catalysis,Shovkovy2013,Tada2020,Gorbar2002,catalysis-original}.
The ZLL is distinct from other LLs in monolayer graphene as it is simultaneously valley and sublattice 
polarized.  There have been numerous suggestions for broken symmetry phases that can cause splitting of the
ZLL and give rise to a $\nu = 0$ quantum Hall 
effect~\cite{dassarma-yang-macdonald,moessner,fuchs,nomura-macdonald,herbut-qhe,herbut2007, 
Jung2009,semenoff-zhou,gusynin-miransky-PRB,barlas-review,kharitonov,Roy2014}.  In Ref.~\cite{Roy2014}, two of
us argued that chiral symmetry breaking orders, i.e. orders that break the sublattice symmetry (e.g. antiferromagnetism or charge 
density wave orders), are likely to be favoured when
one considers the effect of ordering on all filled LLs, not just the ZLL.  Subsequently, the importance of considering multiple
filled LLs was also emphasised in Refs.~\cite{Feshami2016,Lukose2016}.
Such symmetry breaking orders can occur for electrons on a honeycomb lattice for sufficiently strong short-range interactions \cite{HerbutPRL2006,HJR2009},
however, in graphene the strength of these interactions are not sufficient to induce order in the absence of a magnetic 
field~\cite{katsnelson}. By solving mean field gap equations that include the mixing of the filled LLs (also known as LL mixing) 
when chiral symmetry breaking orders are present, we obtained an excellent fit of the excitation gap as a function of perpendicular magnetic
field~\cite{Roy2014} obtained by several different experimental groups~\cite{YacobyPRB2013,NovoselovPNAS,Young2012}.

There are a number of terms in the Hamiltonian that give rise to orders that compete to give the ground state in the $\nu = 0$ state.
Antiferromagnetism can arise from short range Hubbard interactions~\cite{HerbutPRL2006,HJR2009}, and competes with 
ferromagnetic ordering arising from the Zeeman coupling of the magnetic field to spin.  The antiferromagnetic
order is controlled by the magnetic field perpendicular to the graphene sheet, while the Zeeman 
coupling scales with the total magnetic field.  Hence, it is to be expected that increasing the total 
field at fixed perpendicular magnetic field should lead to a transition from an antiferromagnetic
to a ferromagnetic state \cite{kharitonov}.

The competition between different states can be affected by the finite extent and 
temperature of the sample.  In the case of either an antiferromagnet or a ferromagnet, both phases 
are gapped in the bulk, but can be distinguished by their edge states -- a purely ferromagnetic state 
in the ZLL of graphene has gapless edge modes giving Hall conductivity $\sigma_{xy}=2 e^2/h$ \cite{abanin-edge,FertigBrey2006}, 
whereas an easy-plane antiferromagnet has gapped edge states.   There have been several transport experiments on 
graphene in a tilted field~\cite{Young2012,Young2014}, which have demonstrated that the 
edge conductance in the $\nu=0$ state changes from $G = 0$ to $G \simeq 2e^2/h$ with increasing
parallel magnetic field~\cite{Young2014}, and this has been interpreted as a transition from an antiferromagnetic state to a 
ferromagnetic state. These considerations have spurred theoretical investigations of edge states for the 
$\nu = 0$ quantum Hall state~\cite{Jung2009,kharitonov,Lado2014,Murthy2014,Knothe2015,HuangCazalilla2015,Piyatkovskiy2014,Tikhonov2016,Murthy2016,miransky-edge,Gusynin2009}. 
We now present a summary of our main findings.

\subsection{Summary of Results}

The presence of an edge will generically affect the spatial profile of the order parameter near the edge. 
Studies of $\nu = 0$ quantum Hall edges have either calculated edge states using bulk order parameters \cite{kharitonov,Piyatkovskiy2014}
or allowed for the spatial variation of order parameters in the vicinity of the 
edge \cite{Jung2009,Lado2014,Murthy2016,Tikhonov2016,Knothe2015,Murthy2014,HuangCazalilla2015}. The relationship between ordering in the
bulk and ordering in the vicinity of the edge has not yet been quantitatively compared with experiment.
In this paper we extend the approach used to obtain quantitative agreement with bulk measurements in 
Ref.~\cite{Roy2014} and apply it to consider measurements of edge transport reported by Young {\it et al.} \cite{Young2014}. In particular, we use the magnetic field dependence for the bulk gaps obtained in Ref.~\cite{Roy2014} as 
input for calculations of edge states.  We first calculate edge states ignoring spatial variation of the order
parameter, and determine the behaviour of the states and gaps as a function of tilted field (shown in Fig.~\ref{fig:edgegap_hard}).  These results are 
in qualitative but not quantitative agreement with experiment, motivating us to consider the effect of
spatial variations of the order parameters in the presence of an edge.  We include these spatial variations
phenomenologically, using a profile for the order parameters based on the results of Ref.~\cite{Lado2014},
and find that for a graphene flake on a substrate placed in a perpendicular field $B_\perp = 0.7$ T, the gap at the edge closes for a parallel
field of $B_\parallel \sim 40$ T, in reasonable agreement with experiment \cite{Young2014}.
Therefore, chiral symmetry breaking orderings within the framework of magnetic catalysis provide a 
good description of both the bulk and the edge of the $\nu = 0$ quantum Hall state.

We also consider thermal corrections to the gap equations solved in Ref.~\cite{Roy2014}.  This allows us
to obtain estimates for the critical temperature for the $\nu = 0$ and $|\nu| = 1$ quantum Hall states.  
Our estimates are comparable with experimental observations, with the transition in the $\nu = 0$ state 
taking place at about ten times higher temperature scales than for the $|\nu| = 1$ states.  We obtain the 
scaling of the critical temperature with magnetic field and distance to the zero field critical point (shown 
in Figs.~\ref{fig:nu0Tcs} and \ref{fig:nu1Tcs}), and find
that the exponent of the magnetic field dependence appears to have a simple relation to the distance to the 
zero field critical point. Overall, the scaling of the transition temperature ($T_c$) with the 
magnetic field follows closely that of the corresponding chiral symmetry breaking mass at zero temperature~\cite{roy-scaling, roy-inhomogeneous-catalysis}. In particular, $T_c$
respectively scales linearly and sublinearly for weak and intermediate subcritical interaction strengths, while for the zero field critical interaction strength $T_c \sim \sqrt{B}$.

\subsection{Organization}

This paper is structured as follows. In Sec.~\ref{sec:hard_edge} we review the solution of edge states obtained 
using bulk values of the order parameters and show numerical results based on values appropriate to fit the 
results of experiments on the bulk.  In Sec.~\ref{sec:self_consistent} we calculate edge states 
allowing for spatial variation of the order parameters and in Sec.~\ref{sec:thermal} we
consider the effects of thermal fluctuations on the bulk gaps.
Finally, in Sec.~\ref{sec:discussion} we discuss our results and conclude.

\section{Edge States}~\label{sec:hard_edge}

In this section we briefly review the low energy theory of graphene in a strong magnetic field and the 
calculation of edge states in the presence of in-plane antiferromagnetic and easy-axis ferromagnetic order parameters.
These orders can arise due to the presence of short range interactions between electrons \cite{HerbutPRL2006,HJR2009}.
We consider the order parameters to be spatially uniform (a condition that will be relaxed in Sec.~\ref{sec:self_consistent}) 
and study their evolution under a tilted magnetic field using experimentally relevant parameter values.  
If the spatial variation of the order parameters is sufficiently weak that their value close to the edge is 
similar to their bulk value, then this should be a good approximation. 
We present these results as a point of reference for more careful comparison with experiment.

\subsection{Model}

The low energy theory of monolayer graphene can be constructed from fermions residing in the valleys centred on the 
two inequivalent Dirac points $\pm\, \bvec{K}$ at the corners of the Brillouin zone.  The states may be written using an 
eight component spinor $\Psi = [\Psi_\uparrow,\Psi_\downarrow]^T$, where 
$\Psi_\sigma^T = [u_\sigma(+\bvec{K} + \bvec{q}), v_\sigma(+\bvec{K}+\bvec{q}), u_\sigma(-\bvec{K} + \bvec{q}), v_\sigma(-\bvec{K}+\bvec{q})],$
with $|\bvec{q}|\ll |\bvec{K}|$ and $u_\sigma$ and $v_\sigma$ are fermionic annihilation operators on the two sublattices of the 
honeycomb lattice and $\sigma =$ $ \uparrow$ or $\downarrow$ labels electron spin. The effective Dirac dispersion applies out to an ultraviolet momentum cutoff 
$\Lambda$ which is of order $1/a$, where $a$ is the lattice spacing.

%%%%%%%%%%%%%%%%%%%%%%%%%%%%%%%%%%%%%%%%%%%%%%%%%%%%%
%%%%%%%%%%%%%%%%%%%%%%%%%%%%%%%%%%%%%%%%%%%%%%%%%%%%%
%%%%%%%%%%%%%%%%%%%%%%%%%%%%%%%%%%%%%%%%%%%%%%%%%%%%%
\begin{figure}[t!]
 \includegraphics[width=7cm]{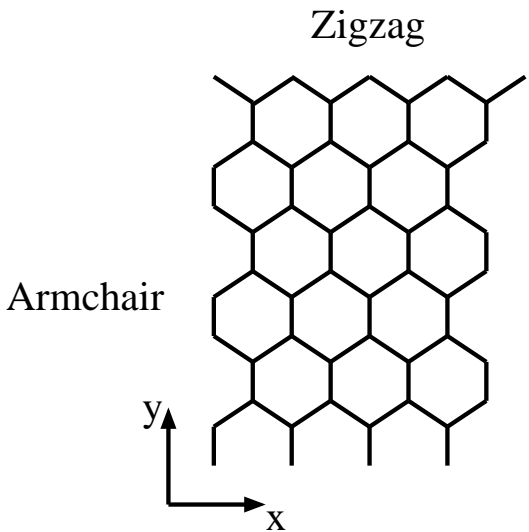}
 \caption{Boundary conditions for a finite graphene sheet, with zigzag and armchair edges illustrated.}
\label{fig:boundaryconds}
\end{figure}
%%%%%%%%%%%%%%%%%%%%%%%%%%%%%%%%%%%%%%%%%%%%%%%%%%%%%
%%%%%%%%%%%%%%%%%%%%%%%%%%%%%%%%%%%%%%%%%%%%%%%%%%%%%
%%%%%%%%%%%%%%%%%%%%%%%%%%%%%%%%%%%%%%%%%%%%%%%%%%%%%

In this basis the Hamiltonian has the 
structure ${\rm spin} \otimes {\rm valley} \otimes {\rm sublattice}$ and allowing for both antiferromagnetic and ferromagnetic 
ordering the Hamiltonian takes the form \cite{herbut2007}
$$ H = I_2 \otimes H_0 - \left(\bvec{N}\cdot \mbox{\boldmath$\sigma$}\right) \otimes \gamma_0 + \left(\lambda + m_3\right) \sigma_3 \otimes I_4,$$
where $H_0 = i\gamma_0\gamma_j \left(-i\partial_j - A_j\right)$ (using the Einstein summation convention), and the gamma matrices take the form 
$\gamma_0   =  I_2 \otimes \sigma_3$, $\gamma_1  = \sigma_3 \otimes \sigma_2$, $\gamma_2  =  - I_2 \otimes \sigma_1$,
$\gamma_3  =  \sigma_1 \otimes \sigma_2$, and $\gamma_5   =  \sigma_2 \otimes \sigma_2$, where the $\sigma_i$ are the usual 
Pauli matrices.  We use units with $e$, $\hbar$ and $v_F$ set to unity unless otherwise specified.  The parameter $\lambda = g\mu_B B$ is the Zeeman 
coupling and $\bvec{N}$ and $\bvec{m}$ are the N\'{e}el and ferromagnetic order parameters respectively.  These order parameters
arise from representing the short range part of electron-electron interactions with a repulsive on-site Hubbard term
\begin{equation}
H_U = \frac{U}{2} \sum_i n_{i\uparrow} n_{i_\downarrow},
\end{equation}	
	and decomposing it with a mean-field approximation \cite{herbut2007,Roy2014}.  We focus on these
two orders as they appear to be the most relevant for the $\nu = 0$ quantum Hall state \cite{kharitonov,Roy2014}.  Similarly to Ref.~\cite{herbut2007}
we exchange the valley and spin indices so that the Hamiltonian is block diagonal in the valley index: 
\begin{equation} 
	H = H_+ \oplus H_-,
\end{equation}
where $H_\pm$ refers to the $\pm K$ valley and
$H_+$ and $H_-$ are related to the Hamiltonian 
   \begin{eqnarray}
	   {\mathcal  H} & = & H_0 +  i\gamma_0 \gamma_3 N_1 + i \gamma_0 \gamma_5 N_2 + i\gamma_3 \gamma_5 (\lambda + m) ,
    \end{eqnarray}
    via unitary transformations.  We define $N_1$ and $N_2$ as the $x$ and $y$ components of the antiferromagnetic order parameter and $m$ as
    the magnitude of the ferromagnetic order parameter.
    Specifically, $H_+ = U_1^\dagger{\mathcal H} U_1$ where $U_1 = I_2 \oplus i\sigma_2$ and $H_- =U_2^\dagger{\mathcal H} U_2$ where $U_2 = i\sigma_2 \oplus I_2$.
    The eight component spinor is transformed to  $\Psi = [\Psi_+, \Psi_-]^T$, where 
    $\Psi_\pm = [u_\uparrow(\pm\bvec{K}+\bvec{q}), v_\uparrow(\pm\bvec{K}+\bvec{q}), u_\downarrow(\pm\bvec{K}+\bvec{q}), v_\downarrow(\pm\bvec{K}+\bvec{q})]^T$.

We follow a similar approach to calculating the edge states to Pyatkovskiy and Miransky \cite{Piyatkovskiy2014}.
We consider a half-plane in which boundary conditions are imposed on one edge and the condition of normalizability is also applied.
We focus on armchair edges which are illustrated along with zigzag edges in Fig.~\ref{fig:boundaryconds}, and details of our calculations are
provided in Appendix~\ref{app:edge}.

The spectrum can be written as 
\begin{equation}
 E_\sigma = \pm\sqrt{{\mathcal N}_\perp^2 +\left[(\tilde{\lambda}+\tilde{m}) + \sigma \sqrt{- \Omega - 1}\right]^2}, \label{eq:llphysical}
\end{equation}
where ${\mathcal N}_\perp^2 = (N_1^2 + N_2^2)/B$, $\tilde{\lambda} = \lambda/\sqrt{B}$, $\tilde{m} = m/\sqrt{B}$, $\sigma = \pm 1$,
and $\Omega$ can be found by solving an eigenvalue equation for parabolic cylinder functions that 
describe the edge states for the appropriate
boundary condition.  The details of these solutions are discussed in Appendix~\ref{app:edge}, 
and in the limit of an infinite sheet Eq.~\eqref{eq:llphysical}  reduces to
 \begin{equation}
	 E_\sigma = \pm \sqrt{{\mathcal N}_\perp^2 + \left[(\tilde{\lambda}+\tilde{m}) +\sigma \sqrt{2n}\right]^2}, 
 \end{equation}
in agreement with the expected bulk expression \cite{Roy2014}.
The edge states for zigzag boundary conditions are similar, but have some differences from 
those for armchair boundary conditions.  Specifically, for zigzag boundary conditions there
are zero energy dispersionless states, which are not present for armchair boundary 
conditions.   We now present numerical results for the edge states.

\subsection{Numerical Results}~\label{sec:edgenum}

As noted above, several other authors \cite{kharitonov,Piyatkovskiy2014,Gusynin2009} have previously obtained the eigenvalue spectrum in the presence of 
edges assuming a uniform order parameter.  In the work here we test whether this can be done in a quantitative manner or not by utilizing the work of 
Roy {\it et al.} \cite{Roy2014}, in which it was found that by solving two mean field gap equations with two adjustable parameters, quantitative agreement 
could be obtained between measurements of the gap as a function of perpendicular magnetic field for both suspended graphene and graphene on a substrate.  
In particular, we use order parameters obtained by solving the gap equation for the bulk as in Ref.~\cite{Roy2014} as input to the eigenvalue equation Eq.~(\ref{eq:llphysical}).  
We then study the effect of a tilted magnetic field on the spectrum, focusing particularly on the gap at the edge, mirroring the situation in experiments 
described in Ref.~\cite{Young2014}.

In Appendix~\ref{app:gapeqns} we briefly review the formalism for self-consistent gap equations in the bulk. 
The total gap for the $\nu=0$ Hall state in the presence of canted antiferromagnetic (CAF) order is $\Delta_0=\sqrt{N^2_\perp+(\lambda+m)^2}$,
where $N_\perp^2 = N_1^2 + N_2^2$.  The gap equations are solved numerically to find  $N_\perp$, $m$ and $\Delta_0$
as a function of magnetic field as the parameters $\delta_a$ and $\delta_f$ are varied. Results are presented in terms of the 
cutoff energy scale $E_c = \hbar v_F \Lambda \simeq 1$ eV. Physically, $\delta_a$ is the 
distance between the critical coupling for AFM order and the actual value of the coupling and $\delta_f$ is the 
dimensionless coupling for FM order. Explicit expressions for $\delta_a$ and $\delta_f$ are presented in Appendix~\ref{app:gapeqns}. 
A positive value of $\delta_{a}$ corresponds to a subcritical coupling.  We note that there was an 
error in the reported value of $\delta_f$ obtained in fits to experimental data in Ref.~\cite{Roy2014} which 
does not affect other conclusions in that work \cite{footnote_Roy} as the ferromagnetic order has minimal
impact of the overall quality of the fit in a perpendicular magnetic field.

\subsubsection{Edge states in a parallel magnetic field}
\label{sec:edgeBfield}

The size of the bulk gap obtained from the self-consistent approach 
depends on the nature of the substrate, with smaller $\delta_a$ values (and a larger gap) 
for suspended graphene than for graphene placed on a substrate where screening increases $\delta_a$ \cite{Roy2014} and decreases the gap.  
Experimentally the transition from antiferromagnetism to ferromagnetism
is realized by applying a magnetic field parallel to the graphene sheet. 

\begin{figure}[t!]
%	        \includegraphics[width=8cm]{figure2a.eps}
%
% \vspace*{1cm}
%
%	        \includegraphics[width=8cm]{figure2b.eps}
	     \includegraphics[width=8cm]{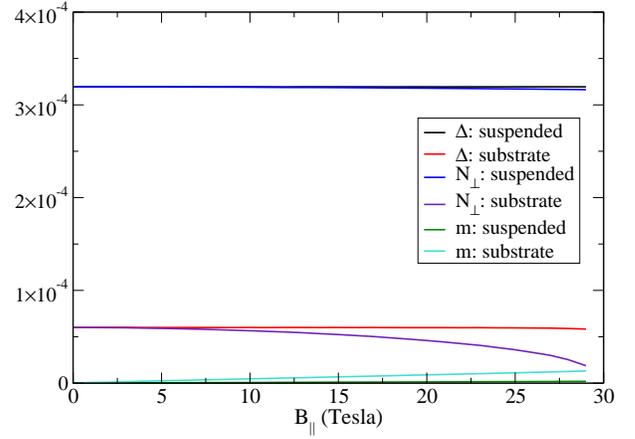}
	\caption{Order parameters in a tilted field for $B_\perp = 0.14$ T for a suspended
	sample ($\delta_a = 0.035$) and a sample on a substrate ($\delta_a = 0.225$).  In both cases
the order parameters are assumed to be spatially uniform in the graphene sheet and  $\delta_f = 1.0$.
Here $\Delta$, $N_\perp$ and $m$ are measured in units of $E_c = \hbar v_F \Lambda$. }
 \label{fig:orderparams_bulk}
\end{figure}

\begin{figure}[t!]
	\includegraphics[width=8cm]{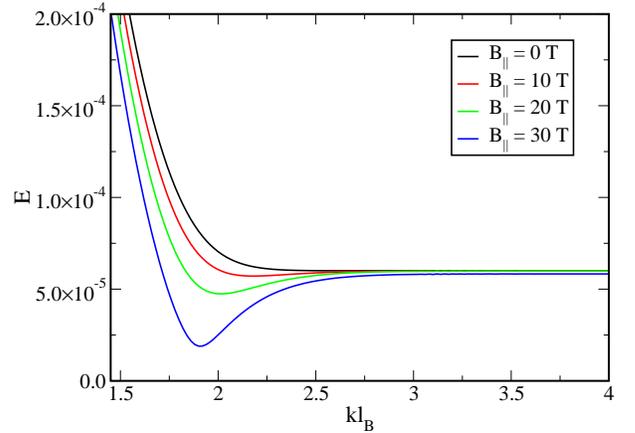}
	\caption{Energy of edge states in units of $E_c$ as a function of $kl_B$ for $B_\perp = 0.14$ T at several different values of parallel
	magnetic field for an armchair boundary condition and parameters suitable for graphene
	on a substrate (i.e. the same as Fig.~\ref{fig:orderparams_bulk} (b)).  The order parameters are 
assumed to be spatially uniform in the graphene sheet.}
\label{fig:edgestates_armchair}
\end{figure}

In Fig.~\ref{fig:orderparams_bulk} we show the total gap ($\Delta$), ferromagnetic order parameter ($m$) 
and antiferromagnetic order parameter ($N_\perp$) as a function of parallel magnetic field for a perpendicular 
magnetic field of  $B_\perp = 0.14$ T for $\delta_a$ and $\delta_f$ values corresponding 
to suspended graphene \cite{Yacoby2012} and graphene on a substrate \cite{Young2012}. 
We see that graphene on a substrate is susceptible to ferromagnetism at much smaller
values of $B_\parallel$ than suspended graphene at the same value of $B_\perp$.  In the experiments in 
Ref.~\cite{Young2012}, the graphene was placed on boron nitride (BN), so for Figs.~\ref{fig:edgestates_armchair}
onward we only show results with $\delta_a = 0.225$, $\delta_f = 1.0$, which are representative parameter values 
for graphene on a BN substrate~\cite{Roy2014}.

%%%%%%%%%%%%%%%%%%%%%%%%%%%%%%%%%%%%%%%%%%%%%%%%%%%%%
%%%%%%%%%%%%%%%%%%%%%%%%%%%%%%%%%%%%%%%%%%%%%%%%%%%%%
%%%%%%%%%%%%%%%%%%%%%%%%%%%%%%%%%%%%%%%%%%%%%%%%%%%%%
%%%%%%%%%%%%%%%%%%%%%%%%%%%%%%%%%%%%%%%%%%%%%%%%%%%%%
%%%%%%%%%%%%%%%%%%%%%%%%%%%%%%%%%%%%%%%%%%%%%%%%%%%%%
%%%%%%%%%%%%%%%%%%%%%%%%%%%%%%%%%%%%%%%%%%%%%%%%%%%%%

%%%%%%%%%%%%%%%%%%%%%%%%%%%%%%%%%%%%%%%%%%%%%%%%%%%%%
%%%%%%%%%%%%%%%%%%%%%%%%%%%%%%%%%%%%%%%%%%%%%%%%%%%%%
%%%%%%%%%%%%%%%%%%%%%%%%%%%%%%%%%%%%%%%%%%%%%%%%%%%%%
%%%%%%%%%%%%%%%%%%%%%%%%%%%%%%%%%%%%%%%%%%%%%%%%%%%%%
%%%%%%%%%%%%%%%%%%%%%%%%%%%%%%%%%%%%%%%%%%%%%%%%%%%%%
%%%%%%%%%%%%%%%%%%%%%%%%%%%%%%%%%%%%%%%%%%%%%%%%%%%%%

%%%%%%%%%%%%%%%%%%%%%%%%%%%%%%%%%%%%%%%%%%%%%%%%%%%%%
%%%%%%%%%%%%%%%%%%%%%%%%%%%%%%%%%%%%%%%%%%%%%%%%%%%%%
%%%%%%%%%%%%%%%%%%%%%%%%%%%%%%%%%%%%%%%%%%%%%%%%%%%%%
\begin{figure}[h]

%	  \vspace*{0.9cm}

   \includegraphics[width=8cm]{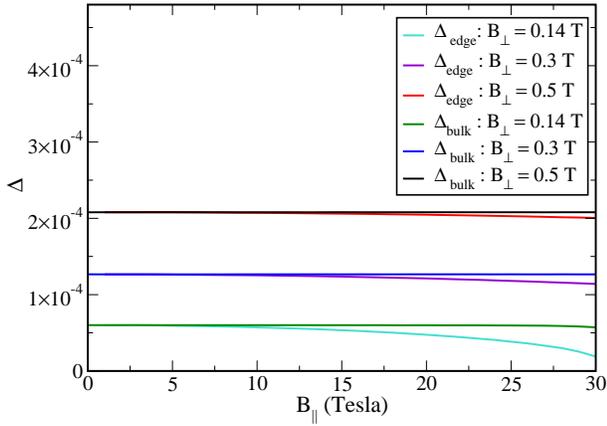}
	\caption{Comparison of the edge gap ($\Delta_{\rm edge}$) and bulk gap ($\Delta_{\rm bulk}$) as a function of parallel magnetic field for $B_\perp = 0.14$ T, 0.3 T and 0.5 T,
	for an armchair edge and parameters appropriate for a substrate (i.e. the same as Fig.~\ref{fig:orderparams_bulk} (b)). The gaps are measured in units of $E_c$.}
\label{fig:edgegap_hard}
\end{figure}
%%%%%%%%%%%%%%%%%%%%%%%%%%%%%%%%%%%%%%%%%%%%%%%%%%%%%
%%%%%%%%%%%%%%%%%%%%%%%%%%%%%%%%%%%%%%%%%%%%%%%%%%%%%
%%%%%%%%%%%%%%%%%%%%%%%%%%%%%%%%%%%%%%%%%%%%%%%%%%%%%

As detailed in Appendix~\ref{app:edge}, we use the Landau gauge $\bvec{A} = (0,Bx)$ for armchair boundary
conditions and take $\Psi(x,y) = e^{iky}\Psi(x)$.  In Fig.~\ref{fig:edgestates_armchair} we show the spectrum 
of edge states for armchair boundary conditions 
when $B_\perp = 0.14$ T and for parallel fields ranging from $B_\parallel$ = 0 to 30~T. The energy 
of the edge states is plotted as a function of $kl_B$, where $l_B = \sqrt{\hbar/eB}$ is the
magnetic length and in the infinite system size limit, $kl_B$ corresponds to the centre of the 
Gaussian part of $\Psi(x)$. One can see that for a parallel field of 30 T, the gap at the 
edge is substantially reduced relative
to the bulk.  

The evolution of the edge gap, $\Delta_{\rm edge}$, (and its relation to the bulk gap, $\Delta_{\rm bulk}$) in the $\nu = 0$ state 
for graphene on a substrate for several different values of
$B_\perp$ is shown in Fig.~\ref{fig:edgegap_hard}, and demonstrates that the required field scale for $B_\parallel$ to quench
antiferromagnetism is much larger than 30 T for $B_\perp \gtrsim 0.3$ T.

The behaviour captured in Figs.~\ref{fig:orderparams_bulk} to \ref{fig:edgegap_hard} is in 
good {\em qualitative} agreement with the experimental results obtained by Young {\it et al.} \cite{Young2014},
but not in good {\em quantitative} agreement.  In Ref.~\onlinecite{Young2014}, a field scale of $B_\parallel \sim
35$ T was required to obtain saturation of the conductance at around $G \simeq 1.8 e^2/h$ for a sample with $B_\perp = 1.4$ T
on a BN substrate. This is suggestive of a transition to ferromagnetism at this field scale.  
Assuming the bulk values of the order parameters all the way out to the edge leads us to require a perpendicular field
scale about 10 times smaller than experiment to see the same closing of the gap at the edge.  This suggests that 
the size of the gap at the edge is being overestimated, and that assuming spatially uniform order parameters is 
an oversimplification.

\section{Edge states for spatially varying order parameters}~\label{sec:self_consistent}

%%%%%%%%%%%%%%%%%%%%%%%%%%%%%%%%%%%%%%%%%%%%%%%%%%%%%
%%%%%%%%%%%%%%%%%%%%%%%%%%%%%%%%%%%%%%%%%%%%%%%%%%%%%
%%%%%%%%%%%%%%%%%%%%%%%%%%%%%%%%%%%%%%%%%%%%%%%%%%%%%
\begin{figure}[t!]

           \vspace*{1cm}

	   \includegraphics[width=8cm]{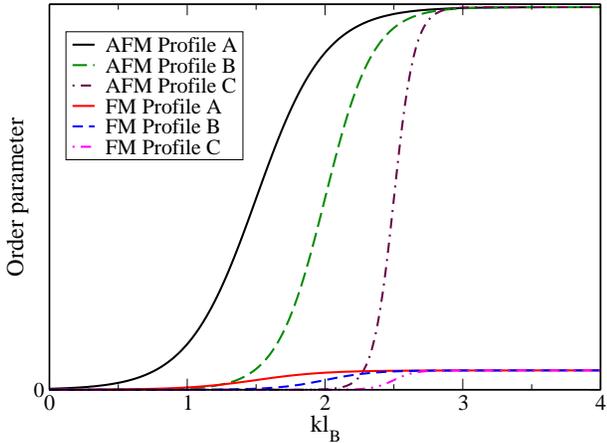}
	   \caption{Phenomenological profiles of the antiferromagnetic (AFM) and ferromagnetic (FM) order parameters for 
	   an armchair edge for three different sigmoidal edge profiles.  
           The order parameters are shown to have relative magnitudes appropriate for parameter values $\delta_a = 0.225$, $\delta_f = 1$ and $B_\perp = 0.14$ T.
	   Profile A decays between $kl_B$ = 2 and 3.  Profile B decays 
	   between $kl_B$ = 1 and 3 and Profile C decays between $kl_B$ = 0 and 3.  The order parameters 
	   take their bulk values (see Fig.~\ref{fig:orderparams_bulk}) 
	   for $kl_B \geq 3$. We do not include a vertical scale as we wish to draw attention to the shape rather than the magnitude (which is fixed by the bulk gap) of the profiles. }
	      \label{fig:profile}
\end{figure}
%%%%%%%%%%%%%%%%%%%%%%%%%%%%%%%%%%%%%%%%%%%%%%%%%%%%%
%%%%%%%%%%%%%%%%%%%%%%%%%%%%%%%%%%%%%%%%%%%%%%%%%%%%%
%%%%%%%%%%%%%%%%%%%%%%%%%%%%%%%%%%%%%%%%%%%%%%%%%%%%%

%%%%%%%%%%%%%%%%%%%%%%%%%%%%%%%%%%%%%%%%%%%%%%%%%%%%%
%%%%%%%%%%%%%%%%%%%%%%%%%%%%%%%%%%%%%%%%%%%%%%%%%%%%%
%%%%%%%%%%%%%%%%%%%%%%%%%%%%%%%%%%%%%%%%%%%%%%%%%%%%%
\begin{figure}[t!]

%           \vspace*{1cm}

	   \includegraphics[width=8cm]{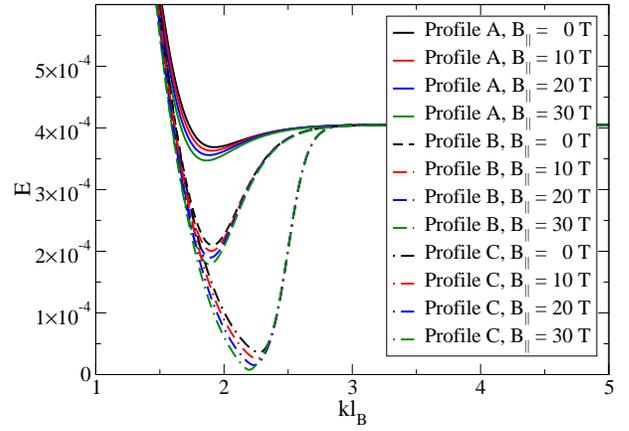}
	      \caption{Spectrum in the presence of spatially varying order parameters for armchair boundary conditions
	      with $B_\perp = 1.4$ T for the profiles A, B, and C introduced in Fig.~\ref{fig:profile}. Energies are measured
	      in units of $E_c$.}
	      \label{fig:renzigzag}
\end{figure}
%%%%%%%%%%%%%%%%%%%%%%%%%%%%%%%%%%%%%%%%%%%%%%%%%%%%%
%%%%%%%%%%%%%%%%%%%%%%%%%%%%%%%%%%%%%%%%%%%%%%%%%%%%%
%%%%%%%%%%%%%%%%%%%%%%%%%%%%%%%%%%%%%%%%%%%%%%%%%%%%%

Having seen in Sec.~\ref{sec:edgeBfield}  that assuming that the order parameters are uniform in the bulk gives 
a qualitatively but not quantitatively correct description of edge states in a parallel magnetic field, we 
generalize our discussion to allow the spatial dependence of order parameters in the presence of an edge.
We do this by allowing $N$ and $m$ to be functions of $k$, i.e. $N_k$ and $m_k$, in addition to $\Omega$ 
which is already a function of $k$.  In general, finding a self-consistent solution for $N_k$ and $m_k$ 
by using a similar approach to the one we used for the bulk is a very challenging problem.  We expect that 
the general behaviour of both $N$ and $m$ is that they will decay from their bulk value in the vicinity of
the edge.  One could envisage generalizing the gap equation approach we use for the bulk by allowing 
spatial variation of order parameters.  This leads to a situation where at each value of $k$, one needs
to self-consistently solve for both the order parameters and $\Omega_k$, which involves performing 
sums over many filled states (which are more complicated in their energy dispersion than Landau levels),
and also carefully devising an appropriate regularization procedure. 
Given that a spatially uniform order parameter profile already produced a qualitatively correct picture
that is compatible with experiment, as a first step towards quantitative agreement, we only account for 
spatial variations of the order parameters phenomenologically.
We make a ``local density approximation'', in which we write the energy eigenvalues for a given $k$ as 
\begin{eqnarray}
	E_{n\bvec{s}}(k) = \pm \sqrt{{{\mathcal N}_\perp}_k^2 + \left[(\tilde{\lambda} + \tilde{m}_k) +\sigma \sqrt{-\Omega_{n\bvec{s}}(k) - 1}\right]^2}. \nonumber \\
\label{eq:EdgeLDA}
\end{eqnarray}
Rather than explicitly solving for ${N_\perp}_k$ and $m_k$, we assume that they have a spatial profile of the form determined
by Lado and Fern\'{a}ndez-Rossier \cite{Lado2014} for an armchair edge, with the bulk value set 
by solving the mean field gap equations.  We allow for three different spatial profiles for the order parameters, A, B and C,
shown in Fig.~\ref{fig:profile},  with A having the slowest
drop-off of the order parameters near the edge through to C having the fastest drop-off of the order parameters near
the edge.  We do not consider zigzag edges, since the order parameters near the edge
are predicted to diverge by Lado and Fern\'{a}ndez-Rossier \cite{Lado2014}, making a phenomenological spatial profile
more difficult to realize.

%%%%%%%%%%%%%%%%%%%%%%%%%%%%%%%%%%%%%%%%%%%%%%%%%%%%%
%%%%%%%%%%%%%%%%%%%%%%%%%%%%%%%%%%%%%%%%%%%%%%%%%%%%%
%%%%%%%%%%%%%%%%%%%%%%%%%%%%%%%%%%%%%%%%%%%%%%%%%%%%%
\begin{figure}[t!]
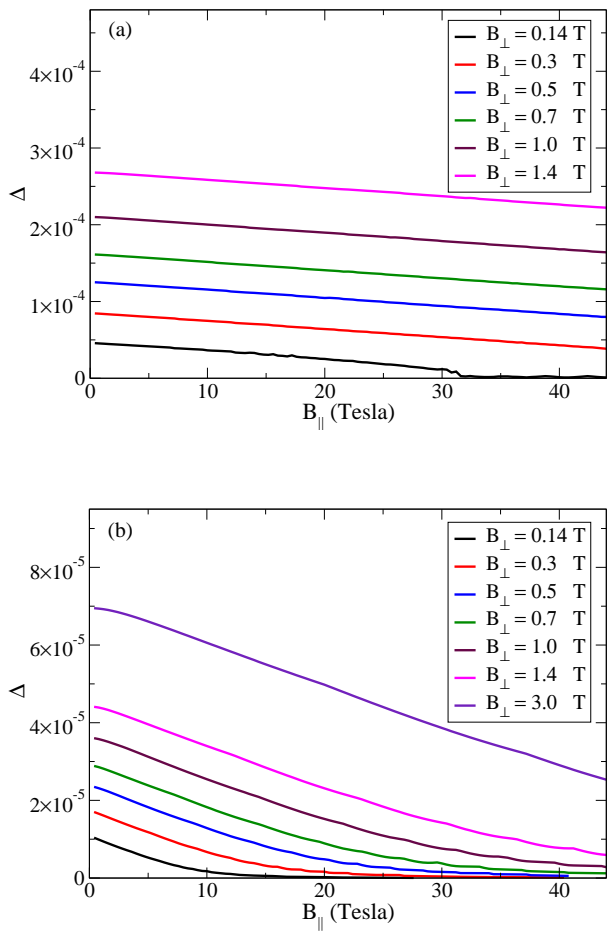

                   \includegraphics[width=8cm]{figure7a.eps}
 
	              \vspace*{1cm}

	           \includegraphics[width=8cm]{figure7b.eps}
		                 \caption{Gaps at the edge as a function of parallel field for a variety of 
					 perpendicular magnetic fields for armchair boundary
				 conditions; (a) Profile B, and (b) Profile C (as defined in Fig.~\ref{fig:profile}). $\Delta$ is 
				 measured in units of $E_c$.}
    \label{fig:edgegaps}
\end{figure}
%%%%%%%%%%%%%%%%%%%%%%%%%%%%%%%%%%%%%%%%%%%%%%%%%%%%%
%%%%%%%%%%%%%%%%%%%%%%%%%%%%%%%%%%%%%%%%%%%%%%%%%%%%%
%%%%%%%%%%%%%%%%%%%%%%%%%%%%%%%%%%%%%%%%%%%%%%%%%%%%%

We solve the self-consistent gap equation at each value of $k$ using the given order parameter and hence
find $\Omega$ as a function of $k$, which allows us to determine the energies of the edge states 
using Eq.~(\ref{eq:EdgeLDA}).  The edge state energies for each of profiles A, B, and C introduced in Fig.~\ref{fig:profile} 
are illustrated in Fig.~\ref{fig:renzigzag}.

We compare the field scales for which we find a transition to ferromagnetism in Fig.~\ref{fig:edgegaps}
for profiles B and C. We find that for profile C for an armchair edge, the field scale for the gap to close
is on the order of $B_\parallel \sim 40$ T when $B_\perp = 0.7$ T, which is 
much closer to the experimental field scale of $B_\parallel \sim 35$ T for $B_\perp = 1.4$ T than the 
uniform order parameter case, for which the gap closes for $B_\parallel > 30$ T for $B_\perp = 0.14$ T.  
Comparison between Fig.~\ref{fig:edgegaps} (a) and (b), corresponding 
to order parameter spatial profiles B and C illustrates that the field scales at which the gap closes
are very sensitive to the spatial variation of the order parameters near the edge. This suggests that 
using the theory developed in Roy {\it et al.} \cite{Roy2014} to determine
the bulk order parameters and then allowing spatial variation phenomenologically
is consistent with experimental results.

The agreement between the results for armchair edges and experiment is much better than the uniform case, but 
there are a number of factors that can be expected to be relevant
at the edge that we have not included here.  These include long-range Coulomb interactions, disorder,
spin fluctuations and Landau level broadening \cite{Hong2019}.  Nevertheless, the development of chiral 
symmetry breaking orders within the framework of magnetic catalysis appears to provide a consistent 
explanation for the experimental observations in both the bulk \cite{Roy2014} and the edge of the 
system for the $\nu = 0$ quantum Hall state.

\section{Thermal corrections}~\label{sec:thermal}

In addition to measurements of the conductance in a magnetic field at fixed temperature, in the 
supplementary materials of Ref.~\cite{Young2014} measurements of conductance as a function of temperature
were also presented.  In this section we generalize the theory for self-consistent 
gap equations to finite temperature and then solve for the transition temperature $T_c$ as a function of $B$ and $\delta$ 
for both $\nu = 0$ and $|\nu| = 1$ states.

\subsection{Magnetic Catalysis at finite temperature}

The zero temperature theory for the gap equations in the magnetic catalysis scenario is reviewed
in Appendix~\ref{app:gapeqns}.  The problem of magnetic catalysis at finite temperature has not been
treated for graphene to our knowledge, but magnetic catalysis in quantum electrodynamics (QED) has been considered in the context of high energy physics~\cite{Klimenko1992,Alexandre2001}. Finite temperature leads to additional terms in the free energy to account for 
entropy that lead to extra terms in the gap equations.

\subsubsection{Gap equations for $\nu = 0$}~\label{sec:gapnu0finiteT}

To formulate the theory of magnetic catalysis at finite temperature, we note that we can write the 
dimensionless free energy ($f$) in the presence of antiferromagnetism and ferromagnetism as

\begin{widetext}
\begin{eqnarray}
f & = & \frac{N^2}{4\lambda_a} + \frac{M^2}{4\lambda_f} - b\left[E_0 + \sum_{\sigma = \pm} \sum_{n=1}^\infty E_{n,\sigma}\right]
	-2bt \left[\ln\left(1 + e^{-\frac{E_0}{t}}\right) + \sum_{\sigma =\pm} \sum_{n=1}^\infty \ln\left(1 + e^{-\frac{E_{n,\sigma}}{t}}\right)\right], 
	\nonumber  \\
  & = & f_0 -2bt \left[\ln\left(1 + e^{-\frac{E_0}{t}}\right) + \sum_{\sigma =\pm} \sum_{n=1}^\infty \ln\left(1 + e^{-\frac{E_{n,\sigma}}{t}}\right)\right],
	\label{eq:finteTNf0}
\end{eqnarray}
where $f = F/hv_F\Lambda^3$, $N = N_\perp/(\hbar v_F \Lambda)$, $M = m/(\hbar v_F \Lambda)$, $\lambda_a = g_a \Lambda/(2\pi\hbar v_F)$, 
$\lambda_f = g_f\Lambda/(2\pi \hbar v_F)$, $t = k_B T/(\hbar v_F \Lambda)$ and $b = h\Lambda^2/(2B)$, with 
$E_0 = \sqrt{N^2 + M^2}$ and $E_{n,\sigma} = \sqrt{N^2 + \left[\sqrt{2nb} + \sigma M\right]^2}$, and $f_0$ is the zero temperature
dimensionless free energy.
Minimizing the free energy gives the gap equations, which may be written in the form
\begin{eqnarray}
\delta_a - Nyf_1^a(x,y) + \frac{N}{\sqrt{\pi}}\left[f_2^a(x,y) - yf_3^a(x)\right] + 2Ny \psi_a(x,y,N,t) & = & 0  ,  \label{eq:finiteTgapN}\\
\frac{m}{N}\delta_f - Nyf^m(x,y) + 2Ny \psi_m(x,y,N,t) & = & 0 , \label{eq:finiteTgapM}
\end{eqnarray}
where $y=B/N^2_\perp$, $x=(\lambda+m)/N_\perp$ and $\delta_a$, $\delta_f$,
$f^a_1(x,y)$, $f^a_2(x,y)$, $f^a_3(x)$ and $f^m(x,y)$ are defined in Appendix~\ref{app:gapeqns}.  The new functions that enter
into the gap equations when thermal effects are included are
\begin{eqnarray}
	\psi_a(x,y,N,t) & = &  \frac{1}{\sqrt{1+x^2}} \frac{1}{1 + \exp\left[\frac{N}{t}\sqrt{1+x^2}\right]} + 
	\sum_{\sigma = \pm} \sum_{n=1}^\infty \frac{1}{\sqrt{1 + (\sqrt{2ny} + \sigma x)^2}} \frac{1}{1 + \exp\left[\frac{N}{t}\sqrt{1 + 
	(\sqrt{2ny} + \sigma x)^2}\right]},  \nonumber \\ & & 
\end{eqnarray}
and
\begin{eqnarray}
	\psi_m(x,y,N,t) & = &  \frac{x}{\sqrt{1+x^2}} \frac{1}{1 + \exp\left[\frac{N}{t}\sqrt{1+x^2}\right]} + 
	\sum_{\sigma = \pm} \sum_{n=1}^\infty \frac{\sigma (\sqrt{2ny} + \sigma x)}{\sqrt{1 + (\sqrt{2ny} + \sigma x)^2}} \frac{1}{1 + \exp\left[\frac{N}{t}\sqrt{1 + 
	(\sqrt{2ny} + \sigma x)^2}\right]} . \nonumber \\ & & 
\end{eqnarray}
The thermal corrections to the gap equations for the in-plane antiferromagnet and the easy-axis ferromagnet are accounted
for in the functions $\psi_a$ and $\psi_m$, respectively.

\subsubsection{Gap equation for $|\nu| = 1$}~\label{sec:gapnu1finiteT}

Roy {\it et al.}~\cite{Roy2014} have argued that the quantum Hall states at $|\nu| = 1$ can be mainly understood
as arising due to charge density wave order (another example of chiral symmetry breaking order on the honeycomb lattice \cite{HJR2009}).  
Hence we can start with the dimensionless free energy ($f$) for $|\nu| = 1$ states, including thermal
contributions
\begin{equation}
		f= \frac{C^2}{4\lambda_c} -b\left[\frac{E_0}{2} + 2 \sum_{n=1}^\infty E_{n}\right]
    -2bt \left[\frac{1}{2}\ln\left(1 + e^{-\frac{E_0}{t}}\right) + 2 \sum_{n=1}^\infty \ln\left(1 + e^{-\frac{E_{n}}{t}}\right)\right], 
\end{equation}
	where $\lambda_c$ is a coupling constant proportional to the nearest neighbour repulsive interaction ($V_1$), 
and obtain a gap equation as before
\begin{equation}
\sqrt{\pi}\delta_c + C\int_0^\infty \frac{ds}{s^\frac{3}{2}}\left[ 1 - \frac{sy e^{-s}}{\tanh(sy)} + \frac{sye^{-s}}{2}\right] + \sqrt{\pi} \psi_c(b,C,t) = 0,
\label{eq:finiteTgapC}
\end{equation}
where $y = b/C^2$ and
\begin{equation}
\psi_c(b,C,t) = \frac{b}{C} \frac{1}{1 + e^{\frac{C}{t}}} + 4b \sum_{n=1}^\infty \frac{1}{E_n} \frac{1}{1 + e^{\frac{E_n}{t}}},
\end{equation}
with $\delta_c = \frac{1}{4\lambda_c} - \frac{1}{\sqrt{\pi}}\int_{\Lambda^{-1}}^\infty ds/s^\frac{3}{2}$, and $E_n = \sqrt{C^2 + 2nb}$.
Thermal corrections due to charge density wave order are introduced by the function $\psi_c$. Here $\delta_c$ measures the distance from the zero field critical interaction strength ($\delta_c=0$) for charge density wave ordering. 

\end{widetext}

\subsection{Numerical Results}

We solve the gap equations found in Secs.~\ref{sec:gapnu0finiteT} and \ref{sec:gapnu1finiteT} numerically 
and present our results below for $\nu = 0$ and $|\nu| = 1$.  All order parameters and temperatures are 
expressed in dimensionless units.

%%%%%%%%%%%%%%%%%%%%%%%%%%%%%%%%%%%%%%%%%%%%%%%%%%%%%
%%%%%%%%%%%%%%%%%%%%%%%%%%%%%%%%%%%%%%%%%%%%%%%%%%%%%
%%%%%%%%%%%%%%%%%%%%%%%%%%%%%%%%%%%%%%%%%%%%%%%%%%%%%
\begin{figure}[t!]
                   \includegraphics[width=8cm]{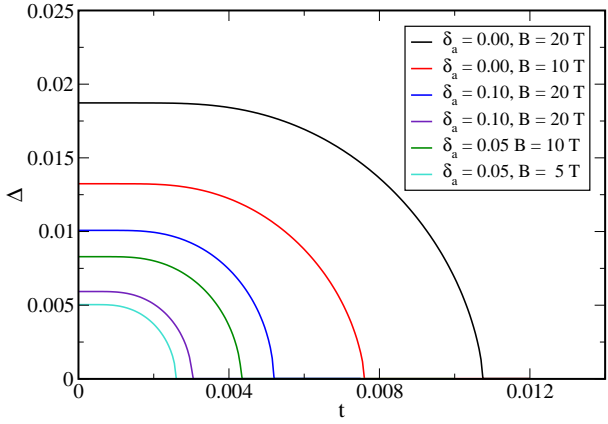}
				 \caption{Gap (in units of $E_c$) for $\nu = 0$ as a function of scaled temperature $t = k_BT/E_c$ for several different field strengths, and several different
				 antiferromagnetic couplings $\delta_a$.  The ferromagnetic coupling is set to $\delta_f = 1$.}
    \label{fig:nu0order}
\end{figure}
%%%%%%%%%%%%%%%%%%%%%%%%%%%%%%%%%%%%%%%%%%%%%%%%%%%%%
%%%%%%%%%%%%%%%%%%%%%%%%%%%%%%%%%%%%%%%%%%%%%%%%%%%%%
%%%%%%%%%%%%%%%%%%%%%%%%%%%%%%%%%%%%%%%%%%%%%%%%%%%%%

%%%%%%%%%%%%%%%%%%%%%%%%%%%%%%%%%%%%%%%%%%%%%%%%%%%%%
%%%%%%%%%%%%%%%%%%%%%%%%%%%%%%%%%%%%%%%%%%%%%%%%%%%%%
%%%%%%%%%%%%%%%%%%%%%%%%%%%%%%%%%%%%%%%%%%%%%%%%%%%%%
\begin{figure}[h]
                   \includegraphics[width=8cm]{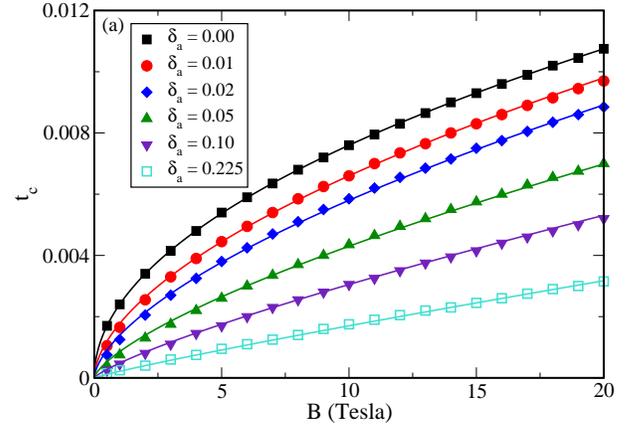}

                       \vspace*{1cm}

		    \includegraphics[width=8cm]{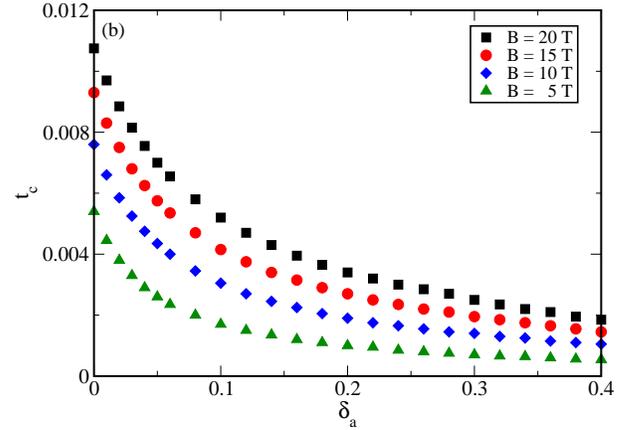}
	       \caption{(a) Critical temperature $t_c = k_B T_c/E_c$ for $\nu = 0$ as a function of magnetic field $B$ for six different values of $\delta_a$. The curves
	       are power law fits to $t_c$ as a function of $B$ using Eqs.~(\ref{eq:TcBpowerlaw}) and (\ref{eq:TcBexponent}).
	       (b) critical temperature $t_c$ for $\nu = 0$ as a function of $\delta_a$ for four different values of magnetic field $B$. 
	           The ferromagnetic coupling is set to $\delta_f = 1$.}
    \label{fig:nu0Tcs}
\end{figure}
%%%%%%%%%%%%%%%%%%%%%%%%%%%%%%%%%%%%%%%%%%%%%%%%%%%%%
%%%%%%%%%%%%%%%%%%%%%%%%%%%%%%%%%%%%%%%%%%%%%%%%%%%%%
%%%%%%%%%%%%%%%%%%%%%%%%%%%%%%%%%%%%%%%%%%%%%%%%%%%%%

\subsubsection{$\nu = 0$}

We solve Eqs.~(\ref{eq:finiteTgapN}) and (\ref{eq:finiteTgapM}) to find the gap as a
function of temperature for given $B_\perp$, $\delta_a$ with fixed $\delta_f=1$.  We consider the situation in which the field is
purely perpendicular to the graphene, so that $N_\perp \gg m$. We show the gap as a function of temperature for a variety of different
coupling strengths and field strengths in Fig.~\ref{fig:nu0order}. Due to the Zeeman coupling the gap $\Delta$ never goes quite to zero, even
when the antiferromagnetic order parameter $N_\perp$ vanishes (we use this to determine the critical temperature $T_c$), but on the scale of Fig.~\ref{fig:nu0order} $N_\perp$ and $\Delta$ are 
indistinguishable.   We distinguish between the dimensionful critical temperature $T_c$ and the dimensionless critical 
temperature $t_c= k_B T_c/E_c$.

The highest $t_c$ value shown in Fig.~\ref{fig:nu0order} of $t \simeq 0.011$ in scaled units corresponds to a physical temperature of $T_c \sim 130$ K.
We extracted the critical temperature $t_c$ as a function of magnetic field $B$ at fixed $\delta_a$ and as a function of $\delta_a$ 
at fixed magnetic field $B$, as illustrated in Fig.~\ref{fig:nu0Tcs}.  We found that we could fit the critical temperature to the following forms.
For fixed $\delta_a$
\begin{equation} 
	t_c \simeq B^\kappa,
\label{eq:TcBpowerlaw}
\end{equation}
with 
\begin{equation}
	\kappa \simeq \frac{1}{2} + \left(\delta_a\right)^{\alpha},
\label{eq:TcBexponent}
\end{equation}
with $\alpha \sim 0.6$,
and for fixed field $B$ and $\delta_a \lesssim 0.15$
\begin{equation}
	t_c \simeq A_B \exp\left[-\left(\frac{\delta_a}{\delta_B}\right)^\beta\right],
\end{equation}
where $A_B$ and $\delta_B$ are field dependent constants and $\beta \simeq 0.8$ for all fields.
Recall that suspended graphene has $\delta_a \simeq 0.035$ and graphene on a substrate has $\delta_a \simeq 0.2$.

\subsubsection{$|\nu| = 1$}
\begin{figure}[t!]
                   \includegraphics[width=8cm]{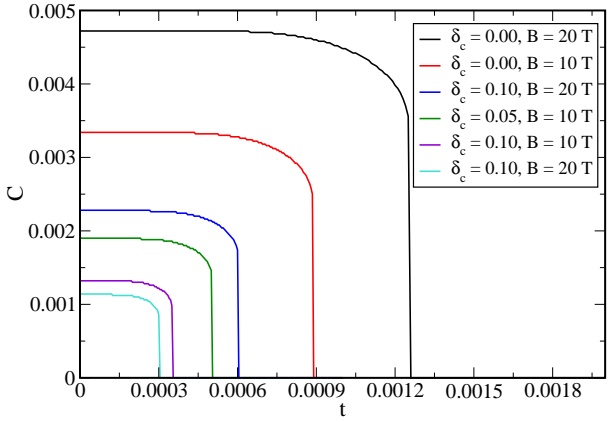}
                                 \caption{Gap for $|\nu| = 1$ as a function of scaled temperature $t$ for several different field strengths, and several different
                                   $\delta_c$.}
    \label{fig:nu1order}
\end{figure}

We obtain the temperature dependence of the CDW gap for $|\nu| = 1$ by solving the gap equation Eq.~(\ref{eq:finiteTgapC}), and display 
the order parameter, $C$, as a function of the dimensionless temperature $t$ for various $B$ and $\delta_c$ in Fig.~\ref{fig:nu1order}.

\begin{figure}[h]
                   \includegraphics[width=8cm]{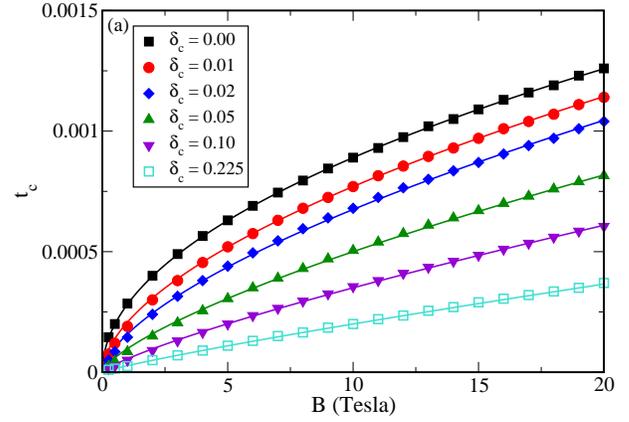}

                     \vspace*{1cm}

                    \includegraphics[width=8cm]{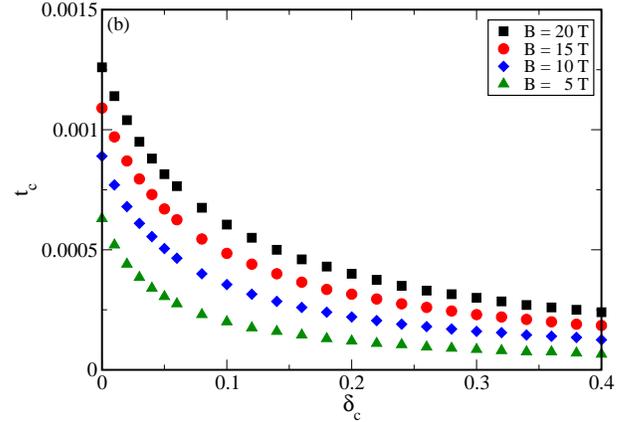} 
	       \caption{(a) Critical temperature $t_c = k_B T_c/E_c$ for $|\nu| = 1$ as a function of magnetic field $B$ for six different values of $\delta_c$. The curves
               are power law fits to $t_c$ as a function of $B$.
	       (b) critical temperature $t_c$ for $|\nu| = 1$ as a function of $\delta_c$ for four different values of magnetic field $B$.}
    \label{fig:nu1Tcs}
\end{figure}

We extracted the critical temperature $t_c$ as a function of magnetic field $B$ at fixed $\delta_c$ and as a function of coupling $\delta_c$
at fixed magnetic field $B$, as illustrated in Fig.~\ref{fig:nu1Tcs}.  We found that we could fit the critical temperature to similar forms that we 
used for $\nu = 0$. For fixed $\delta_c$, $t_c$ follows Eqs.~(\ref{eq:TcBpowerlaw}) and (\ref{eq:TcBexponent}) with $\delta_a$ replaced by $\delta_c$
and $\alpha \sim 0.6$,
and for fixed field and $\delta_c \lesssim 0.15$,
\begin{equation}
	t_c \simeq A_B \exp\left[-\left(\frac{\delta_c}{\delta_B}\right)^\beta\right],
\end{equation}
where $A_B$ and $\delta_B$ are field dependent constants and $\beta \simeq 0.75$ for all fields.

The temperature scale for $|\nu| = 1$ transitions appears to be about an order of magnitude smaller than for 
$\nu = 0$.  This is consistent with the expectation that the zero temperature gap for $|\nu| = 1$ is about an
order of magnitude smaller than the zero temperature gap for $\nu = 0$. 
For both states the scaling of the transition temperature $T_c$ with magnetic field follows closely that
of the associated chiral symmetry breaking mass at zero temperature \cite{roy-scaling,roy-inhomogeneous-catalysis}. In particular, $T_c$ scales linearly and sublinearly for
weak and subcritical interaction strengths and for critical interactions $T_c \sim \sqrt{B}$.

%%%%%%%%%%%%%%%%%%%%%%%%%%%%%%%%%%%%%%%%%%%%%%%%%%%%%
%%%%%%%%%%%%%%%%%%%%%%%%%%%%%%%%%%%%%%%%%%%%%%%%%%%%%
%%%%%%%%%%%%%%%%%%%%%%%%%%%%%%%%%%%%%%%%%%%%%%%%%%%%%
%%%%%%%%%%%%%%%%%%%%%%%%%%%%%%%%%%%%%%%%%%%%%%%%%%%%%
%%%%%%%%%%%%%%%%%%%%%%%%%%%%%%%%%%%%%%%%%%%%%%%%%%%%%
%%%%%%%%%%%%%%%%%%%%%%%%%%%%%%%%%%%%%%%%%%%%%%%%%%%%%

%%%%%%%%%%%%%%%%%%%%%%%%%%%%%%%%%%%%%%%%%%%%%%%%%%%%%
%%%%%%%%%%%%%%%%%%%%%%%%%%%%%%%%%%%%%%%%%%%%%%%%%%%%%
%%%%%%%%%%%%%%%%%%%%%%%%%%%%%%%%%%%%%%%%%%%%%%%%%%%%%
%%%%%%%%%%%%%%%%%%%%%%%%%%%%%%%%%%%%%%%%%%%%%%%%%%%%%
%%%%%%%%%%%%%%%%%%%%%%%%%%%%%%%%%%%%%%%%%%%%%%%%%%%%%
%%%%%%%%%%%%%%%%%%%%%%%%%%%%%%%%%%%%%%%%%%%%%%%%%%%%%

\subsection{Nature of the transition}
We expect that for a single chiral symmetry breaking order parameter, the finite temperature
phase transition should be second order.  For $|\nu| = 1$, 
the transition appears to be second order, with a very steep decline of the charge density wave 
order parameter near $T_c$, consistent with earlier theoretical work
for the Gross-Neveu model \cite{Klimenko1992}. 
 On the other hand, for $\nu = 0$, when there is 
both in-plane antiferromagnetism and easy-axis ferromagnetism, as shown in Fig.~\ref{fig:nu0order}, 
for $\delta_a$ near zero, the transition appears as though it is second order.  We have noted 
that truncation errors in the numerical evaluation of the integrals and sums in the gap 
equations for $\nu = 0$ (e.g. if insufficient Landau levels are included) tend to make the transition 
appear first order \cite{footnote}. 

\begin{figure}[t!]
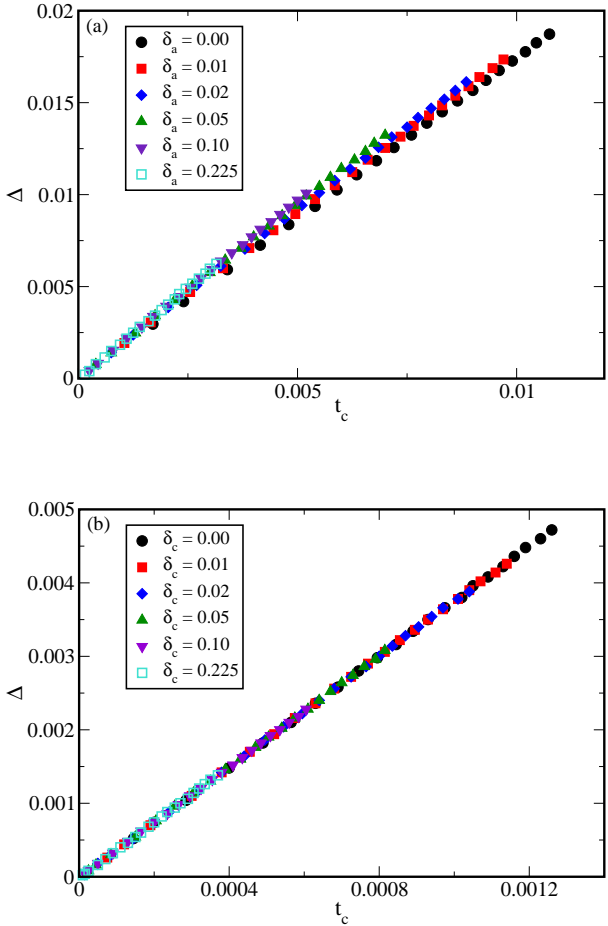

                   \includegraphics[width=8cm]{nu0gap.eps}

                        \vspace*{1cm}

                    \includegraphics[width=8cm]{nu1gap.eps}
      \caption{Zero temperature gap, $\Delta$, against critical temperature $t_c$ for various coupling strengths for 
	(a) $\nu = 0$ with only antiferromagnet ordering (ferromagnetism turned off) and (b) $|\nu| = 1$ with charge-density-wave ordering. 
	For $\nu=1$ we find $\Delta/t_c \approx 3.75$ for a wide range of interaction strengths ($\delta_c$).}
    \label{fig:gaps}
\end{figure}

We compared the zero temperature gap, $\Delta$, to $t_c$ (both in dimensionless units) for the different
field and coupling strengths considered above.  For $\nu = 0$ we find that the 
relationship between $\Delta$ and $t_c$ is linear, and ranges from 
$\Delta /t_c = 1.75$ at $\delta_a = 0$ to $\Delta/t_c = 2$ for $\delta_a = 0.225$, as illustrated 
in Fig.~\ref{fig:gaps} (a).  For $|\nu| = 1$ on the other hand, we find the universal scaling that
$\Delta/t_c \approx 3.75$ for any $\delta_c$ as illustrated in Fig.~\ref{fig:gaps} (b).

%%%%%%%%%%%%%%%%%%%%%%%%%%%%%%%%%%%%%%%%%%%%%%%%%%%%%
%%%%%%%%%%%%%%%%%%%%%%%%%%%%%%%%%%%%%%%%%%%%%%%%%%%%%
%%%%%%%%%%%%%%%%%%%%%%%%%%%%%%%%%%%%%%%%%%%%%%%%%%%%%
%%%%%%%%%%%%%%%%%%%%%%%%%%%%%%%%%%%%%%%%%%%%%%%%%%%%%
%%%%%%%%%%%%%%%%%%%%%%%%%%%%%%%%%%%%%%%%%%%%%%%%%%%%%
%%%%%%%%%%%%%%%%%%%%%%%%%%%%%%%%%%%%%%%%%%%%%%%%%%%%%

\section{Discussion and Conclusions}~\label{sec:discussion}

In Ref.~\cite{Roy2014} it was shown that the gaps in the bulk for the $\nu = 0$ and $|\nu| = 1$
quantum Hall states in monolayer graphene can be explained with a picture based on chiral symmetry
breaking. While the $\nu = 0$ state is compatible with a canted antiferromagnet, the $|\nu| = 1$ states
are likely to be due to charge density wave ordering.  Both in-plane antiferromagnet and charge 
density wave orderings are examples of chiral symmetry breaking orders on the honeycomb lattice.
In this work we explored the effects of finite sample size and finite temperature within 
the same scenario.  We focused on edge states at $\nu = 0$ and the field and interaction strength 
dependence of $T_c$ for both $\nu = 0$ and $|\nu| = 1$.  

Experiments \cite{Young2014} have suggested that in the $\nu = 0$ state  there is a transition 
from antiferromagnetism to ferromagnetism if a strong enough parallel field is applied
at fixed perpendicular field based on obervations of the increase in conductance with tilted field.
This can be understood as the increasing strength of ferromagnetism
decreasing the gap at the edge, and consequently allowing edge transport.   Using the theory for
the bulk order parameters discussed in Ref.~\cite{Roy2014} we show that tilted fields have much 
more effect for graphene samples on a substrate than for suspended graphene.  We study the edge states
assuming that the order parameter at the edge is the same as in the bulk and find qualitative but not 
quantitative agreement with experiment in that the gap decreases with increasing tilted field, but at 
a much smaller perpendicular field. However, when we allow spatial variation of the order parameters near
the edge with a phenomenological profile based on work in Ref.~\cite{Lado2014}, the tilted field scale 
at which the edge gap closes is $B_\parallel \sim 40 $ T for $B_\perp = 0.7$ T, as illustrated 
in Fig.~\ref{fig:edgegaps}, which is in a similar
range to experiment, emphasizing the importance of spatial variation of the order parameter near 
edges \cite{Jung2009,Tikhonov2016,HuangCazalilla2015,Murthy2016,Lado2014,Murthy2014,Knothe2015}.

The effect of increasing temperature on both the $\nu = 0$ and $|\nu| = 1$ states is to give rise to a 
transition to a disordered state at a non-zero critical temperature $T_c$ as illustrated in Figs.~\ref{fig:nu0order} and
\ref{fig:nu1order}.  We calculated $T_c$ as a function of magnetic field, $B$, and 
distance to the critical point, $\delta$, in Figs.~\ref{fig:nu0Tcs} and \ref{fig:nu1Tcs}.  We found that the 
functional form of the dependence of $T_c$ on magnetic field is the same for both $\nu = 0$
and $|\nu| = 1$ and observe similar behaviour between the two states for the dependence of $T_c$ on the distance to the critical
point.  The magnetic field dependence of $T_c$ can be tested experimentally and potentially used as a way
to extract $\delta_a$ or $\delta_c$ for a given sample.  To study the $\delta$ dependence of $T_c$, using
gated samples and changing the strength of screening is a way to change the distance from the critical point.

It should be noted that we have only considered the effects of short range interactions.  In graphene there will also
be effects from long range Coulomb interactions.  These can affect the bulk behaviour and edge reconstruction 
and appear to be needed to obtain a full understanding of the dependence of the gap on field for $|\nu| = 1$ states \cite{Roy2014}.  

Our results for $\nu = 0$ edge states are consistent with chiral symmetry breaking in the zeroth Landau level of monolayer graphene
giving rise to the $\nu = 0$ quantum Hall state.  Recent consideration of fractional quantum Hall states of graphene
has led to the suggestion that chiral symmetry breaking may be a unifying feature of quantum Hall states in the 
zeroth Landau level of monolayer graphene \cite{Narayanan2018}.  Measurement of the critical temperature of 
the $\nu = 0$ and $|\nu| = 1$ integer quantum Hall states, particularly focusing on the scaling of $T_c$ with 
magnetic field would be an important additional test of this 
scenario and we look forward to the results of experiments investigating this behaviour.

\acknowledgements

M.R.C.F., S.N. and M.P.K. were supported by NSERC.  M.P.K. thanks the Max Planck Institute for the 
Physics of Complex Systems in Dresden for hospitality while a portion of this work was completed.
B. R. was supported by a Startup Grant from Lehigh University.

\begin{appendix}

	\section{Calculation of edge states}~\label{app:edge}
		
In this appendix we give a brief discussion of the calculation of the edge state eigenvalues in graphene
	for the cases of armchair and zigzag boundary conditions (see Fig.~\ref{fig:boundaryconds}).
The boundary conditions arising for zigzag and armchair graphene edges are, respectively~\cite{Piyatkovskiy2014}
(for $H$ block diagonal in the valley index)
\begin{eqnarray}
	        \left(I_8 + I_2 \otimes I_2 \otimes \sigma_3\right) \Psi(y=0) = 0 \label{eq:zigzag} , \\
		        \left(I_8 + I_2 \otimes I_2 \otimes \sigma_1\right) \Psi(x=0) = 0 \label{eq:armchair} .
		\end{eqnarray}
		Assuming spatial uniformity of the order parameters,
		the eigenvalue problem can be solved analytically for the zigzag edge by utilizing the Landau gauge $\textbf{A} = (-By,0)$
		and letting $\Psi(x,y) = \exp(i k x)\Psi(y)$.  For armchair boundary conditions we take
		$\textbf{A} = (0,Bx)$ and $\Psi(x,y) = \exp(i k y)\Psi(x)$.

		After the valley degree of freedom has been extracted, write the spinors in the $4 \times 4$ representation used for ${\mathcal H}$
		as $\psi = \left[\psi_1,\psi_2,\psi_3,\psi_4\right]^{T}$ and the valley spinors $\Psi_\pm$ can then be written as
		\begin{eqnarray}
			        \Psi_{+} = \left[\psi_1,\psi_2,-\psi_4,\psi_3\right]^{T}, \label{eq:upvalleyspinor} \\
				        \Psi_{-} = \left[-\psi_2,\psi_1,\psi_3,\psi_4\right]^{T}. \label{eq:downvalleyspinor}
				\end{eqnarray}

				Defining $ \xi = \frac{k_x + By}{\sqrt{B}}$, which implies $\partial_2 = \sqrt{B} \partial_\xi$, we can write the
				eigenvalue equation ${\mathcal H}\psi = \varepsilon\psi$ as
				\begin{eqnarray}
					        -E_+ \psi_1 - \partial_- \psi_2 + {\mathcal N}_- \psi_4 & = & 0 \label{eq:eqI}  , \\
					        \partial_+ \psi_1  - E_+ \psi_2 + {\mathcal N}_- \psi_3 & = & 0 \label{eq:eqII} , \\
					        {\mathcal N}_+ \psi_2 - E_- \psi_3 - \partial_+ \psi_4 & = & 0 \label{eq:eqIII} , \\
					        {\mathcal N}_+ \psi_1 + \partial_- \psi_3 - E_- \psi_4 & = & 0 \label{eq:eqIV}  .
				\end{eqnarray}
				where we introduced $\partial_\pm = \partial_\xi \pm \xi$ and ${\mathcal N}_\pm = (N_1 \pm i N_2)/\sqrt{B}$ and
				$E_\pm = (\varepsilon \pm (\lambda + m))/\sqrt{B}$.

				Focusing on the $\psi_1$ component of the spinor, we obtain the eigenvalue equation
				\begin{equation}
					        \left[\partial_- \partial_+ - 1\right] \psi_1(\xi) = \Omega \psi_1(\xi), 
						\label{eq:parabolic}
					\end{equation}
		where the eigenvalues $\Omega$ are related to the energy eigenvalues by
		\begin{equation}
		        E = \pm\sqrt{{\mathcal N}_\perp^2 +\left[(\tilde{\lambda}+\tilde{m}) \pm\sqrt{- \Omega - 1}\right]^2}, \label{eq:llphysicalapp}
		\end{equation}
		where ${\mathcal N}_\perp^2 = {\mathcal N}_+^2 + {\mathcal N}_-^2 - (N_1^2 + N_2^2)/B$ and 
		$\tilde{\lambda} = \lambda/\sqrt{B}$, $\tilde{m} = m/\sqrt{B}$.
		Equation~(\ref{eq:parabolic}) has solutions which are parabolic cylinder functions with eigenvalue $\Omega$,
		which, when the solution is required to be normalizable as $y \to -\infty$ has the solutions: \cite{miransky-edge,Piyatkovskiy2014,Stegun}

\begin{equation}
\psi_{i} = g_i\begin{cases} U\left(\frac{1}{2}\Omega,\sqrt{2}\,\xi\right) &;  i = 1, 4 \\ U\left(\frac{1}{2}\Omega + 1,\sqrt{2}\,\xi\right) &;  i = 2,3, \end{cases} \nonumber
\end{equation}
where $U(a,z)$ is the even parabolic cylinder function \cite{Stegun}
and the proportionality constants $g_i$, are defined by
	\begin{eqnarray}
       g_{1} &=& 1, \nonumber \\
      g_2 &=& -\frac{N_\perp^2 - \left(\Omega + 1\right) - \varepsilon_-\varepsilon_{+}}{2\sqrt{2}\left(\lambda+m\right)}, \nonumber \\
   g_{3} &=& \frac{1}{N_{-}}\left[\varepsilon_+ g_{2}+\frac{1}{\sqrt{2}}\left(\Omega + 1\right)\right], \nonumber \\
 g_{4} &=& \frac{1}{N_{-}}\left[-\sqrt{2}g_{2}+\varepsilon_{+}\right], 
	\end{eqnarray}
where $\varepsilon_\pm = \varepsilon \pm (\lambda+m)$.

 \subsubsection{Zigzag edge}

 The zigzag boundary condition Eq.~(\ref{eq:zigzag}) gives the following constraints on the spinor $\Psi_{+}$
 \begin{eqnarray}
         0 &=& U\left(\frac{1}{2}\Omega,\sqrt{2}\,\xi_0\right), \label{eq:zigzagvalleyup1} \\
	 0 &=& \operatorname{Re}\{g_4\}\,U\left(\frac{1}{2}\Omega,\sqrt{2}\,\xi_0\right), \label{eq:zigzagvalleyup2}
\end{eqnarray}
and for the spinor $\Psi_{-}$
\begin{eqnarray}
       0 &=& \operatorname{Re}\{g_2\}\,U\left(\frac{1}{2}\Omega + 1,\sqrt{2}\,\xi_0\right), \label{eq:zigzagvalleydown1}  \\
 0 &=& \operatorname{Re}\{g_3\}\,U\left(\frac{1}{2}\Omega + 1,\sqrt{2}\,\xi_0\right), \label{eq:zigzagvalleydown2}
 \end{eqnarray}
where $\xi_0 = k/\sqrt{B}$. Only the real part of the coefficients $g_i$ enter the boundary conditions since the energies
 $\varepsilon$, and hence the eigenvalues $\Omega$, are required to be real valued.

 These boundary conditions fix the eigenvalues $\Omega = \Omega_{n\sigma s}$, where $n \in \mathbb{Z}_{\geq 0}$ labels the ``branch''
 index and $\sigma = \pm$ labels the spin and $s = \pm$ labels the sublattice degree of freedom.

 An important feature of the eigenvalue equations \eqref{eq:zigzagvalleyup1}-\eqref{eq:zigzagvalleydown2} is that they all take the form
 of a constant times a parabolic cylinder function, meaning that the eigenvalue determined by the vanishing of the momentum-independent
 coefficients $g_i(\Lambda) = 0$, here labeled by a branch index of $n = 0$, is dispersionless, while all of the higher branches
 (which determine the higher LLs via Eq.~\eqref{eq:llphysical}) are determined by the vanishing of the parabolic cylinder
 function in question. 
 This is in contrast to the armchair
	 edge discussed below, where the eigenvalue is determined by a linear combination of parabolic cylinder functions, which
	 prevents one from factoring out the zero mode edge state.

 Neglecting the $n = 0$ eigenvalue, the higher eigenvalues with $n \in \mathbb{N}$ then become independent of the spin index and are completely determined by the equations
 \begin{equation}
 0 = U\left(\frac{1}{2}\Omega,\sqrt{2}\, \xi_0\right),  
\end{equation}
	 and
 \begin{equation}
 0 = U\left(\frac{1}{2}\Omega + 1,\sqrt{2}\,\xi_0\right), 
\end{equation}
 in the $+$ and $-$ valleys, respectively. The roots $\Omega_{ns}$ of these equations are all negative definite as a function of $\xi_0$ \cite{Stegun}.
If one takes the bulk limit, which corresponds to $\xi \rightarrow \infty$,
		 then the eigenvalues $\Omega \rightarrow -2n-1$ where $n \in \mathbb{Z}_{\geq 0}$ \cite{Stegun,Piyatkovskiy2014},
	 and the expression for the energy eigenvalues, Eq.~\eqref{eq:llphysical},  reduces to
 \begin{equation}
        E = \pm \sqrt{{\mathcal N}_\perp^2 + \left[(\tilde{\lambda}+\tilde{m}) \pm \sqrt{2n}\right]^2}, 
 \end{equation}
 in agreement with the expected expression \cite{Roy2014}.

 \subsubsection{Armchair edge}

 The armchair boundary condition Eq.~\eqref{eq:armchair} imposes the pair of constraints
 \begin{eqnarray}
 0 &=& U\left(\frac{1}{2}\Omega,\sqrt{2}\,\xi_0\right) + s'\operatorname{Re}\{g_2\}\, U\left(\frac{1}{2}\Omega +1 ,\sqrt{2}\,\xi_0\right), \nonumber \\
	 & & \label{eq:armchairbc1} \\
 0 &=& \operatorname{Re}\{g_3\}\,U\left(\frac{1}{2}\Omega+1,\sqrt{2}\,\xi_0\right) \nonumber \\
	  &  & - s'\operatorname{Re}\{g_4\} \, U\left(\frac{1}{2}\Omega,\sqrt{2}\,\xi_0\right), \label{eq:armchairbc2}
 \end{eqnarray}
 on the $s' = \pm$ valley.  We solve Eqs.~(\ref{eq:armchairbc1}) and (\ref{eq:armchairbc2}) numerically to obtain $\Omega$, and hence, using Eq.~(\ref{eq:llphysicalapp}),
 the  energy spectrum of the edge modes, which are displayed in Secs.~\ref{sec:edgenum} and \ref{sec:self_consistent}.

\section{Zero temperature gap equations}
\label{app:gapeqns}
The gap equations that we use for the bulk have been discussed in considerable detail 
elsewhere \cite{herbut2007,roy-BLG,Roy2014}. We give a brief summary of their derivation
here for $\nu = 0$ in order to facilitate our discussion of magnetic catalysis at finite temperature in Sec.~\ref{sec:thermal}.
In the bulk, when there are both AFM and FM orders the LLs have the form 
$\pm E_{n,\sigma}$, where \cite{herbut2007}
\begin{equation}\label{DiracLL}
	        E_{n,\sigma}=\sqrt{N^2_\perp+[(N^2_3 + 2 n B)^{1/2} + \sigma (m+\lambda)]^2},
	\end{equation}
with $N_3$ and $N_\perp$ the easy-axis and easy-plane components of the Neel order parameter respectively and $\sigma=\pm$ the
	two spin projections. The degeneracy of the LLs is $D = 1/(\pi l^2_B)$ for $n \geq 1$ 
and $1/(2\pi l^2_B)$ for $n=0$ \cite{Roy2014}.
The corresponding zero temperature free energy \cite{herbut2007, roy-BLG,Roy2014} is obtained from a sum over
filled LLs (for $n\geq 1$):
	\begin{eqnarray}
		        F_0=\frac{N^2_\perp+N^2_3}{4 g_a}+\frac{m^2}{4 g_f}-D \sum_{\sigma=\pm} \left[ \frac{1}{2}E_{0,\sigma} + \sum_{n \geq 1} E_{n,\sigma} \right], \nonumber \\
		\end{eqnarray}
where $g_a$ ($g_f$) are couplings arising from short-range interactions, such as on-site Hubbard repulsion,
that support AFM (FM) order \cite{comment-1,Roy2014}. 
For non-trivial Zeeman coupling ($\lambda \neq 0$), $F_0$ is minimized when $N_3 \equiv 0$ \cite{herbut2007,roy-BLG}.
Therefore, the Zeeman coupling restricts the AFM order to the easy-plane and simultaneously allows
FM order parallel to the magnetic field. 
Taking $N_3 = 0$ and then minimizing $F_0$ with respect to $N_\perp$ and $m$ leads to coupled gap
equations \cite{Roy2014}.
Ferromagnetic order splits all the filled LLs, including the zeroth one, while easy plane AFM order lowers the energy of all 
of the filled LLs in addition to splitting the ZLL. Hence the contribution from the filled LLs with 
$n \geq 1$ in the first (second) gap equation add up (cancel). Consequently, the second gap equation is 
free of divergences, but the first one exhibits an ultraviolet divergence which can be regularized as
discussed in Refs.~\cite{herbut-qhe,roy-scaling,Roy2014} and written in terms of 
 $\delta_a = \frac{1}{4\lambda_a} - \frac{1}{\sqrt{\pi}}\int_{\Lambda^{-1}}^\infty ds/s^\frac{3}{2}$
  where $(\lambda^a_c)^{-1}=\int^\infty_{\Lambda^{-1}} ds/s^{3/2}$ is the zero magnetic field critical onsite interaction
 strength for AFM ordering \cite{HerbutPRL2006,HJR2009,herbut-assaad2013}  and $\delta_f = 1/2\lambda_f$.  The relation between
 $\lambda_{a,f}$ and $g_{a,f}$ was specified in Sec.~\ref{sec:gapnu0finiteT}.
Thus the two gap equations, after regularization, can be written compactly as
\begin{widetext}
 \begin{eqnarray}
\delta_a- N_\perp y f^a_1(x,y) + \frac{N_\perp}{\sqrt{\pi}} \left(f_2^a(x,y) - yf_3^a(x)\right) & = & 0, \\
	 \frac{m}{N_\perp} \delta_f-N_\perp y f^m(x,y) & = & 0, 
 \end{eqnarray}
where we have introduced $y=B/N^2_\perp$ and $x=(\lambda+m)/N_\perp$. 
The various functions appearing in these two equations are given by
 \begin{eqnarray}
 f^a_1(x,y)&=&\sum_{n \geq 0} \sum_{\sigma=\pm} \left[ \frac{1}{\left[ 1+ \left(\sqrt{2 n y} + \sigma x \right)^2 \right]^{1/2}}
             -\frac{1}{\left( 1+ 2 n y\right)^{1/2}} \right],  \\
 f^a_2(x,y) & = & \int^\infty_0 \frac{ds}{s^{3/2}} \left[1-s y e^{-s} \coth(s y) \right] , \\
 f^a_3(x,y)&=&\int^\infty_0 \frac{ds}{s^{1/2}} e^{-s} \left(1-e^{-s x^2} \right),  \\
 f^m(x,y) & = & \left[\sum_{n \geq 0} \sum_{\sigma=\pm} \frac{\sigma \left(\sqrt{2 n y}+\sigma x \right)}{\left[ 1+ 
\left(\sqrt{2 n y} +\sigma x \right)^2\right]^{1/2}}\right]   -\frac{x}{(1+x^2)^{1/2}} . 
\end{eqnarray}
\end{widetext}

\end{appendix}


\begin{thebibliography}{99}


\bibitem{qhe-graphene-1} K. S. Novoselov, A. K. Geim, S. V. Morozov, D. Jiang, M. I. Katsnelson, I. V. Grigorieva, S. V. Dubonos, and A. A. Firsov, Nature (London) {\bf 438}, 197 (2005).

\bibitem{qhe-graphene-2} Y. Zhang, Y.-W. Tan, H. L. Stormer, and P. Kim, Nature (London) {\bf 438}, 201 (2005).

\bibitem{sharapov} V. P. Gusynin, and S. G. Sharapov, Phys. Rev. Lett. {\bf 95}, 146801 (2005).


\bibitem{Zhang2007} Y. Zhang, Z. Jiang, J. P. Small, M. S. Purewal, Y.-W. Tan, M. Fazlollahi, J. D. Chudow, J. A. Jaszczak, H. L. Stormer, and P. Kim, Phys. Rev. Lett. {\bf 96}, 136806 (2006).

\bibitem{Andrei2010} I. Skachko, X. Du, F. Duerr, A. Luican, D. A. Abanin, L. S. Levitov, E.Y. Andrei, Phil. Trans. R. Soc. A {\bf 368}, 5403 (2010).

\bibitem{YacobyPRB2013} D. A. Abanin, B. E. Feldman, A. Yacoby, and B. I. Halperin, Phys. Rev. B {\bf 88}, 115407 (2013).

\bibitem{NovoselovPNAS} G. L. Yu, R. Jalil, B. Belle, A. S. Mayorov, P. Blake, F. Schedin, S. V. Morozov, L. A. Ponomarenko, F. Chiappini, S. Wiedmann, U. Zeitler, M. I. Katsnelson, A. K. Geim, K. S. Novoselov, and D. C. Elias, 
	Proc. Nat. Acad. Sci. {\bf 110}, 3282 (2013).

\bibitem{Young2012}  A. F. Young, C. R. Dean, L. Wang, H. Ren,  P. Cadden-Zimansky, K. Watanabe, T. Taniguchi, J. Hone, K. L. Shepard, and P. Kim, Nat. Phys. {\bf 8}, 550 (2012).

\bibitem{Andrei2009} X. Du, I. Bkachko, F. Duerr, A. Luican, E. Y. Andrei, Nature {\bf 462}, 192 (2009).

\bibitem{Dean2011}  C. R. Dean, A. F. Young, P. Cadden-Zimansky, L. Wang, H. Ren, K. Watanabe, T. Taniguchi, P. Kim, J. Hone, and K. L. Shepard, Nat. Phys.
	{\bf 7}, 693 (2011).

\bibitem{Yacoby2012} B. E. Feldman, B. Krauss, J. H. Smet, A. Yacoby, Science {\bf 337}, 1196 (2012).

\bibitem{Li2019} S.-Y. Li, Y. Zhang, L. J. Yin, and L. He, Phys. Rev. B {\bf 100}, 085437 (2019).

\bibitem{Hong2019} S. J. Hong, C. Belke, J. C. Rode, B. Brechtken, and R. J. Haug, electronic preprint arXiv:1908.02420v1.

\bibitem{Roy2014} B. Roy, M. P. Kennett, and S. Das Sarma, Phys. Rev. B {\bf 90}, 201409(R) (2014).

\bibitem{Khveshchenko2001} D. V. Khveshchenko, Phys. Rev. Lett. {\bf 87}, 246802 (2001); H. Leal and D. V. Khveshchenko, Nucl. Phys. {\bf B687}, 323 (2004).

\bibitem{HerbutPRL2006} I. F. Herbut, Phys. Rev. Lett. {\bf 97}, 146401 (2006).
	
\bibitem{HJR2009} I. F. Herbut, V. Juri\v{c}i\'{c}, and B. Roy, Phys. Rev. B {\bf 79}, 085116 (2009).

\bibitem{herbut-qhe} I. F. Herbut, Phys. Rev. B {\bf 75}, 165411 (2007).

\bibitem{herbut2007} I. F. Herbut, Phys. Rev. B {\bf 76}, 085432 (2007).

\bibitem{KunYang2007} K. Yang, Solid State Commun. {\bf 143}, 27 (2007).

\bibitem{semenoff-zhou} G. W. Semenoff and F. Zhou, JHEP {\bf 1107}, 037 (2011).

\bibitem{goerbig-review} M. O. Goerbig, Rev. Mod. Phys. {\bf 83}, 1193 (2011).

\bibitem{barlas-review} Y. Barlas, K. Yang, and A. H. MacDonald, Nanotechnology {\bf 23}, 052001 (2012).

\bibitem{kharitonov} M. Kharitonov, Phys. Rev. B {\bf 85}, 155439 (2012).

\bibitem{Narayanan2018} S. Narayanan, B. Roy, and M. P. Kennett, Phys. Rev. B {\bf 98}, 235411 (2018).

\bibitem{catalysis-original} V. P. Gusynin, V. A. Miransky, and I. A. Shovkovy, Phys. Rev. Lett. {\bf 73}, 3499 (1994).

\bibitem{Gorbar2002} E. V. Gorbar, V. P. Gusynin, V. A. Miransky, I. A. Shovkovy, Phys. Rev. B {\bf 66}, 045108 (2002).

\bibitem{roy-scaling} I. F. Herbut and B. Roy, Phys. Rev. B {\bf 77}, 245438 (2008).

\bibitem{roy-inhomogeneous-catalysis} B. Roy and I. F. Herbut, Phys. Rev. B {\bf 83}, 195422 (2011).

\bibitem{Shovkovy2013} I. A. Shovkovy, {\it Magnetic Catalysis: A Review}. In: Kharzeev D., Landsteiner K., Schmitt A., Yee HU. (eds) 
	Strongly Interacting Matter in Magnetic Fields. Lecture Notes in Physics, vol 871. Springer, Berlin, Heidelberg (2013).

\bibitem{Tada2020} Y. Tada, Phys. Rev. Research {\bf 2}, 033363 (2020).

\bibitem{dassarma-yang-macdonald} K. Yang, S. Das Sarma, and A. H. MacDonald, Phys. Rev. B {\bf 74}, 075423 (2006).

\bibitem{moessner} M. O. Goerbig, R. Moessner, and B. Doucot, Phys. Rev. B {\bf 74}, 161407(R) (2006).

\bibitem{fuchs} J.-N. Fuchs and P. Lederer, Phys. Rev. Lett. {\bf 98}, 016803 (2007).

\bibitem{nomura-macdonald} K. Nomura and A. H. MacDonald, Phys. Rev. Lett. {\bf 96}, 256602 (2006).

\bibitem{Jung2009} J. Jung and A. H. MacDonald, Phys. Rev. B {\bf 80}, 235417 (2009).

\bibitem{gusynin-miransky-PRB} V. P. Gusynin, V. A. Miransky, S. G. Sharapov, and I. A. Shovkovy, Phys. Rev. B {\bf 74}, 195429 (2006).

\bibitem{Feshami2016} B. Feshami and H. A. Fertig, Phys. Rev. B {\bf 94}, 245435 (2016).

\bibitem{Lukose2016} V. Lukose and R. Shankar, Phys. Rev. B {\bf 94}, 085135 (2016).

\bibitem{katsnelson} T. O. Wehling, E. Sasioglu, C. Friedrich, A. I. Lichtenstein, M. I. Katsnelson, and S. Bl\"{u}gel, Phys. Rev. Lett. {\bf 106},  236805 (2011).

\bibitem{abanin-edge} D. A. Abanin, K. S. Novoselov, U. Zeitler, P. A. Lee, A. K. Geim, and L. S. Levitov, Phys. Rev. Lett. {\bf 98}, 196806 (2007).

\bibitem{FertigBrey2006} H. A. Fertig and L. Brey, Phys. Rev. Lett. {\bf 97}, 116805 (2006).

\bibitem{Young2014} A. F. Young, J. D. Sanchez-Yamagishi, B. Hunt, S. H. Choi, K. Watanabe, T. Taniguchi, R. C. Ashoori, and P. Jarillo-Herrero, Nature {\bf 505}, 528 (2014).

\bibitem{Piyatkovskiy2014} P. K. Pyatkovskiy, and V. A. Miransky, Phys. Rev. B {\bf90}, 195407 (2014).

\bibitem{Tikhonov2016} P. Tikhonov, E. Shimshoni, H. A. Fertig, and G. Murthy, Phys. Rev. B {\bf 93}, 115137 (2016).

\bibitem{HuangCazalilla2015} C. Huang and M. A. Cazalilla, Phys. Rev. B {\bf 92}, 155124 (2015).

\bibitem{Murthy2016} G. Murthy, E. Shimshoni, and H. A. Fertig, Phys. Rev. B {\bf 93}, 045105 (2016).

\bibitem{Lado2014} J. L. Lado and J. Fern\'{a}ndez-Rossier, Phys. Rev. B {\bf 90}, 165429 (2014).

\bibitem{Murthy2014} G. Murthy, E. Shimshoni, and H. A. Fertig, Phys. Rev. B {\bf 90}, 241410(R) (2014).

\bibitem{Knothe2015} A. Knothe and T. Jolicoeur, Phys. Rev. B {\bf 92}, 165110 (2015).

\bibitem{miransky-edge} V. P. Gusynin, V. A. Miransky, S. G. Sharapov, and I. A. Shovkovy, Phys. Rev. B {\bf 77}, 205409 (2008).

\bibitem{Gusynin2009} V. P. Gusynin, V. A. Miransky, S. G. Sharapov, I. A. Shovkovy, and C. M. Wyenberg, Phys. Rev. B {\bf 79}, 115431 (2009).

\bibitem{footnote_Roy} In Ref.~\cite{Roy2014}, the coupling $\delta_f$ is reported as 0.05. 
	The correct value of $\delta_f$ for the fits in Ref.~\cite{Roy2014} is $\delta_f = 1.0$.

\bibitem{Klimenko1992} K. G. Klimenko, Z. Phys. C {\bf 54}, 323 (1992).
 
\bibitem{Alexandre2001} J. Alexandre, K. Farakos, and G. Koutsoumbas, Phys. Rev. D {\bf 63}, 065015 (2001).

\bibitem{footnote} If the transition is first order, then this is likely a consequence of the coupling to the non-zero
	ferromagnetic order. The possibility of a first order transition when one has two coupled order parameters can be seen from 
the Landau free energy $$ f = hm + bm^2 + aN^2 + cN^4 + dmN^2,$$
where $m$ is the ferromagnetic order parameter and $N$ the in-plane antiferromagnetic order parameter, which is a minimal
free energy for the $\nu = 0$ state. Here, $a,b,c,d$ are unknown coeffcients, and $h$ can be identified with external magnetic field. The equilibrium value of $m$ is non-zero, and when substituted into $f$ gives a
free energy of the form $$ f = a^\prime N^2 + c^\prime N^4,$$ with $c^\prime = c - d^2/2b$.  If $c^\prime < 0$, then $N^6$
terms are required to stabilize the free energy and a first order phase transition results.

\bibitem{Stegun} M. Abramowitz, and I. A. Stegun, {\it Handbook of Mathematical Functions with Formulas, Graphs, and Mathematical Tables}, U.S. Gov. Printing Office, Washiington, D.C. (1972).

\bibitem{roy-BLG} See B. Roy, Phys. Rev. B {\bf 89}, 201401(R) (2014) for discussion on easy-plane layer AFM in bilayer graphene.

\bibitem{comment-1} Although both $\lambda_a$ and $\lambda_f$ arise from onsite repulsion ($U$), in the presence of magnetic fields $\lambda_a \neq \lambda_f$~\cite{herbut2007, roy-BLG}.

\bibitem{herbut-assaad2013} F. F. Assaad and I. F. Herbut, Phys. Rev. X {\bf 3}, 031010 (2013).

\end{thebibliography}
\end{document}